# Information−Theoretic Free Energy as Emotion Potential

## Emotional Valence as a Function of Complexity and Novelty

Hideyoshi Yanagisawa

The University of Tokyo, Hongo 7-3-1, Bunkyo, Tokyo, Japan
hide@mech.t.u-tokyo.ac.jp

**Abstract.** This study extends the mathematical model of emotion dimensions that we previously proposed to consider perceived complexity as well as novelty, as a source of arousal potential. Berlyne's hedonic function of arousal potential (or the inverse U-shaped curve, the so-called Wundt curve) is assumed. We modeled the arousal potential as information contents to be processed in the brain after sensory stimuli are perceived (or recognized), which we termed sensory surprisal. We mathematically demonstrated that sensory surprisal represents free energy, and it is equivalent to a summation of information gain (or information from novelty) and perceived complexity (or information from complexity), which are the collative variables forming the arousal potential. We demonstrated empirical evidence with visual stimuli (profile shapes of butterfly) supporting the hypothesis that the summation of perceived novelty and complexity shapes the inverse U-shaped beauty function. We discussed the potential of free energy as a mathematical principle explaining emotion initiators.

## 1   Introduction

In our previous study (Yanagisawa, Kawamata, & Ueda, 2019), we proposed a mathematical model of dominant emotion dimensions: arousal (or intensity) and valence (i.e., positivity or negativity) (Lang, 1995; Russell, 1980) associated with novelty. We formalized the arousal with Kullback–Leibler (KL) divergence (Kullback & Leibler, 1951) of Bayesian posterior from the prior, which we termed *information gain*. We confirmed that the information gain corresponds to surprise with participants' responses using event-related potential P300 and subjective reports of surprise to novel stimuli. We considered that the information gain function could be used as a mathematical model explaining the arousal potential of Berlyne's theory (Berlyne, 1970). Berlyne suggested that an appropriate level of arousal potential might induce a positive hedonic response, but an extreme arousal potential might induce negative responses. The hedonic function of the arousal potential shapes the inverse U, the so-called Wundt curve, as shown in Fig. 1.

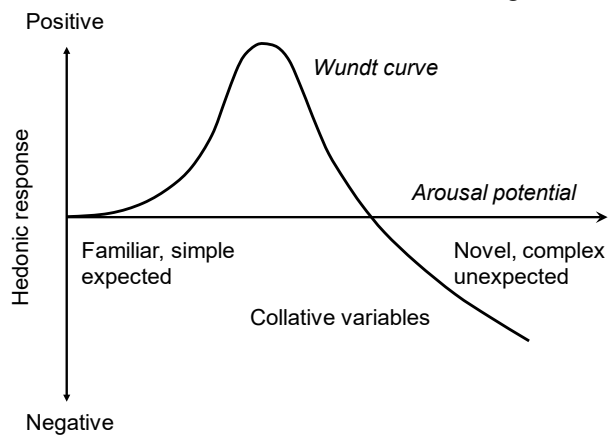

**Fig. 1.** Hedonic function of arousal potential. Collative variables such as novelty and complexity are assumed to be sources of arousal potential.

Novelty is, however, only one of the sources of arousal potential called collative variables. Berlyne exemplified another collative variable such as complexity. Information gain represents information content gained from novelty, but it does not represent information contents regarding complexity. In this study, we updated our arousal model to include complexity. We considered information content to be processed when one perceives sensory stimuli. We mathematically demonstrated that the information content is equivalent to free-energy as defined in physics (original definition), statistics, and more recently, neuroscience (Friston, Kilner, & Harrison, 2006). We revealed that this free energy comprises the summation of information from both perceived novelty and perceived complexity. We then demonstrated that the summation of perceived novelty and complexity can be correlated with the arousal potential from empirical evidence of visual stimuli (profile shapes of a butterfly).

## 2 Free Energy as a Source of Emotional Arousal Potential

### 2.1 Belief Distributions in Perceptions of Sensory Stimuli

The core idea of our model is that the total information content to be processed in the brain after perceiving sensory stimuli represents the potential cognitive load in the brain, and this potential cognitive load works as a source of emotional arousal (i.e. Berlyne's arousal potential) which then works as an initiator of subsequent emotions. According to information theory (Shannon, Weaver, Blahut, & Hajek, 1949), information content is defined by the negative of the log probability, $-\log p$, where $p$ is the probability of an event. Thus, we start to consider belief probability distributions in a situation where one obtains information content by perceiving the external world. Here, we defined *perception* as an estimation of the causes of sensory stimuli. Sensory stimuli are coded to neural activity in the brain, such as the firing rate of certain neuronal populations, through sensory organs (Yanagisawa, 2016). We termed the neural signals *sensory data*. Assume the sensory data as a random variable $X$ and follows certain probability distributions. Now, one obtains $n$ sensory input $X^n = (X_1, ..., X_n)$. True distribution is usually unknown. Instead of the true distributions, we assume that a brain has belief distributions $p(x)$. We can write $p(x)$ as a marginal distribution using the cause of sensory data representing continuous random variables $\theta \in R$.

$$\begin{aligned} p(X^n) &= \int_{-\infty}^{\infty} p(X^n, \theta) d\theta \\ &= \int_{-\infty}^{\infty} p(X^n \mid \theta) p(\theta) d\theta \\ &= \int_{-\infty}^{\infty} \prod_{i=1}^{n} p(X_i \mid \theta) p(\theta) d\theta \end{aligned} \tag{1}$$

Here, we consider that a joint probability distribution between sensory data and its causes $p(x, \theta)$ was learned by past experience of perceiving varied sensory data through one's life. We termed $p(x, \theta)$ *generative model* because it can generate sensory data $x$ from the cause $\theta$. We can decompose this model into the likelihood function $p(X^n \mid \theta)$ and prior $p(\theta)$. The likelihood function represents the likelihood of a cause $\theta$ of the sensory data $X^n$. The prior refers to the belief distributions of a cause $\theta$ before experiencing the sensory data $X^n$. Thus, the belief distributions $p(x)$ are estimated by a product of prior and likelihood.

## 2.2 Sensory Surprisal and Free Energy

Now, we formalize the information content of sensory data $x$ as $-\log p(x)$. We termed the information content *sensory surprisal* because sensory stimuli providing new information evokes surprise. We considered sensory surprisal as information content that a brain processed after $\theta$ is estimated (or perceived) based on incoming sensory data $X^n$. The Bayesian theorem estimates posterior $p(\theta|X^n)$ representing belief distributions of perception of $\theta$ based on the sensory data $X^n$ as in formula (2):

$$p(\theta|X^n) = \frac{p(\theta)p(X^n|\theta)}{p(X^n)} \tag{2}$$

With the Bayesian theorem, we can write sensory surprisal using prior, posterior, and likelihood functions using the following formula:

$$-\log p(X^n) = \log \frac{p(\theta|X^n)}{p(\theta)p(X^n|\theta)} \tag{3}$$

Then, we averaged the right side of formula (3) over the posterior:

$$\left\langle \log \frac{p(\theta|X^n)}{p(\theta)p(X^n|\theta)} \right\rangle_{p(\theta|X^n)} = -\log p(X^n) =: \varphi \tag{4}$$

We define $\varphi$ as *free energy*. We considered that free energy is an information content that brain potentially processes posterior to perceiving or recognizing sensory stimuli.

## 2.3 Free Energy as a Summation of Novelty and Complexity

We can decompose the free energy into two terms using the following formulas:

$$\varphi = -\log p(X^n) = G + U \tag{5}$$

$$G = KL(p(\theta|X^n) \| p(\theta)) \sim \text{perceived novelty} \tag{6}$$

$$U = \left\langle -\log p(X^n|\theta) \right\rangle_{p(\theta|X^n)} \sim \text{perceived complexity} \tag{7}$$

The first term *G*, KL divergence from posterior to prior, represents a gap between prior belief and posterior belief as formula (6), or unexpectedness. We previously defined this term as *information gain* and experimentally confirmed that it corresponds to human surprise induced by unexpected and novel stimuli (Yanagisawa et al., 2019). It also corresponds to *Bayesian surprise* (Itti & Baldi, 2009).

The second term *U* is the negative log-likelihood averaged over the posterior (formula (7)). This term increases as both the variance of the likelihood due to uncertain sensory data and KL divergence from prior to likelihood increases. We can interpret *U* as the complexity or uncertainty regarding a perceived cause of stimuli. Thus, we termed *U perceived complexity*. (It corresponds to *inverse accuracy*.)

In summary, the free energy, representing sensory surprisal averaged over the posterior, is equivalent to a summation of information gain (unexpectedness or novelty) and perceived complexity (or perceived uncertainty).

## 2.4 Relations Between Free Energy Definition in Physics, Bayesian Statistics, and Neuroscience

Free energy has been defined in various scientific disciplines. Historically, Helmholtz originated the concept of free energy in physics (specifically in thermodynamics). Subsequently, statistical physics( or statistical dynamics) derived Helmholtz's Free-energy as a function of inverse temperature $\beta$ and a partition function (or sum over states) $Z(\beta)$ using Boltzmann's entropy:

$$F(\beta) = -\frac{1}{\beta}\log Z(\beta) \tag{8}$$

Bayesian statistics analogically used the free energy formula (8) and defined a partition function $Z(\beta)$ using prior and likelihood functions:

$$Z_n(\beta) = \int_\theta p(\theta)\prod_{i=1}^n p(X_i\mid\theta)^\beta d\theta \tag{9}$$

When $\beta =1$, the free energy in Bayesian statistics is called the marginal likelihood or evidence. This definition corresponds to sensory surprisal and our definition of free energy:

$$F_n(1) = -\log Z(1) = -\log p(X^n) \tag{10}$$

More recently, in the field of neuroscience, Friston et al. introduced the idea of free energy minimization as a principle to explain varied brain activities such as perceptions and actions(Friston et al., 2006). He defined variational free energy *VFE* as KL divergence from the recognition density $q(\theta)$ to generative model $p(\theta, X^n)$:

$$VFE = KL(q(\theta)\parallel p(\theta, X^n)) = KL(q(\theta)\parallel p(\theta\mid X^n)) - \log p(X^n) \tag{11}$$

The first term, KL-divergence, is non-negative by definition. Thus, the second term is the lower limit of the variational free energy.

$$VFE \geq -\log p(X^n) \tag{12}$$

When the recognition density is variationally approximated to posterior, the variational free energy decreases and is close to the lower limit. The lower limit of the variational free energy corresponds to our definition of free energy.

As discussed above, all formulations of free energy in various disciplines are mathematically equivalent, but differ in approach, focus, and philosophy.

## 3 An Empirical Evidence: Beauty of butterfly

### 3.1 Method

We conducted an experiment with participants to verify the hypothesis derived from the model prediction: summation of the perception of novelty and complexity works as an arousal potential, and shapes an inverse-U-shaped hedonic function. We used the profiles of butterflies as visual stimuli. We prepared 48 samples to vary the complexity and familiarity (novelty) of the outline shapes. 20 university students (15 males and 5 females; age range, 20 – 24 years) participated in the experiment. We asked the participants to score the perceived novelty, complexity of shape, and beauty (as a hedonic response) for all samples using a Likert scale of 9 levels for each evaluation item. We used "familiar-unfamiliar", "simple-complex", and "ugly-beautiful" for scales of novelty, complexity and beauty, respectively.

### 3.2 Results and Discussion

We tested the hypothesis that beauty forms an inverse-U-shaped function of the summation of complexity and novelty. We conducted quadratic curve fitting using average scores of novelty, complexity, and beauty obtained from the 20 participants for 48 samples. The results showed a significant quadratic curve relationship between the beauty and summation of novelty and complexity scores. (quadratic estimation: $R^2 = 0.583$, $p < 0.05$; liner estimation: $R^2 = 0.0037$, $p = 0.68$). The quadratic curve was concave down. The estimation formula was $y=-0.25x^2+1.83x+2.75$, as shown in Fig.2. Fig.3 shows the result of the Gaussian curve fitting of the same data as Fig.2. The estimated curve shows a peak around the middle of the score (around 5). These results suggest that the score of beauty is an inverse-U shaped function of summation of novelty and complexity for the sample, and the summation of novelty and complexity work as an arousal potential.

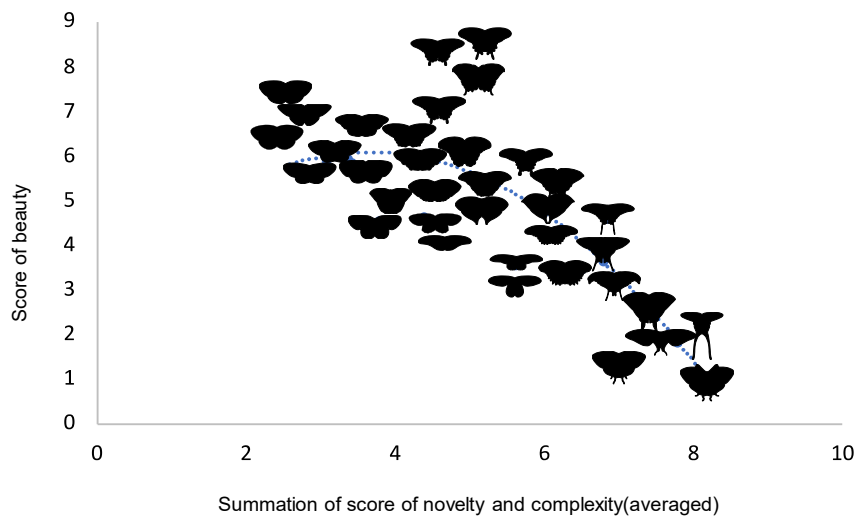

**Fig. 2.** Beauty as a function of summation of novelty and complexity

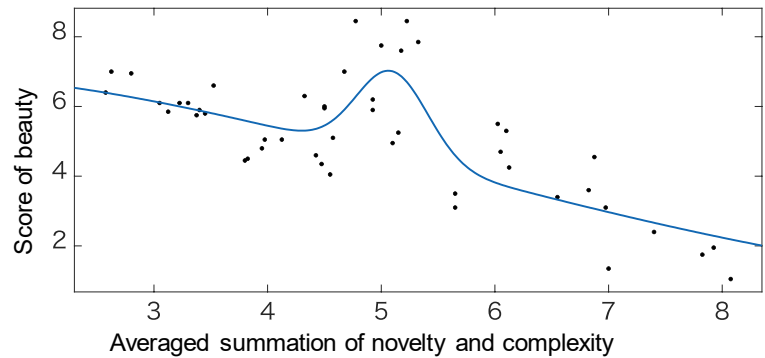

**Fig. 3.** Gaussian curve fitting of beauty as a function of the summation of novelty and complexity

## 4    Concluding Remarks

We mathematically revealed the relations of information contents of perceived sensory stimuli, free energy, and arousal potential (i.e., the primary emotion dimension). Information content to be processed after perception of sensory stimuli, or sensory surprisal, corresponded to formulation of free energy commonly found in varied disciplines such as physics, statistics, and neuroscience. We demonstrated that free energy can be represented as a summation of two terms of information content: information gain and inverse log likelihood averaged over posterior. We considered that the two terms represent novelty and complexity, respectively. These two factors are Berlyne's collative variables, which are sources of arousal potential. Our previous model (Yanagisawa et al., 2019) only considered the first term: information gain (novelty). Thus, free energy is an extension of our previous model to be more general, including the second term regarding perceived complexity. Indeed, several empirical studies have shown that the hedonic function of perceived complexity shows an inverse-U shape (Hung & Chen, 2012; Lévy, MacRae, & Köster, 2006). Mathematical treatments using free energy suggest that the sum of novelty and complexity works as the arousal potential. We demonstrated empirical evidence of the hypothesis using visual stimuli: profile shapes of butterflies. The experimental results showed that beauty is an inverse-U shaped function of the summation of novelty and complexity. Further experimental evidence will be appreciated using various objects, including artifacts as well as natural objects, to ensure validity.

  The human brain is an organ that processes information. Information contents to be processed take a mean cognitive load. The cognitive load consumes biological energy. According to Friston's theory, the brain perceives the causes of sensory data so that the variational free energy is minimized. The free energy minimization of equilibrium is a law in both physical and biological systems. Our definition of free energy corresponds to the minimized variational free energy and focused on information content remaining after perception of sensory stimuli (or recognition) is done where recognition density can be approximated to Bayesian posterior. Our expectation is that the (minimized or remained) free energy is used to activate emotional arousal and subsequent emotions such as valence, and free energy is a general principle of emotion potential. Emotion motivates certain actions such as approach and avoidance. Active inference suggests that free energy can be reduced by acting to gain sensory evidences (Friston et al., 2015). This implies that emotions are a function that initiates action to reduce free energy, and the remaining free energy activates the function.

# Acknowledgments

This study was supported by KAKEN grant number 18H03318 from the Japan Society for the Promotion of Science. We thank Makoto Watarizaki of University of Tokyo for supporting making the empirical evidence.